\documentclass[12pt]{article}
\usepackage{amsmath}
\usepackage{epsfig}
\usepackage{amsfonts}
\usepackage{feynmp}

 \title{Special relativity and reduced spin density matrices}
 \author{Cezary Gonera\thanks{supported by the {\L}\'od\'z University grant N$^0$ 690.},\\ 
 Piotr Kosi\'nski$^*$, Pawe\'l Ma\'slanka$^*$ \\
Department of Theoretical Physics II \\
University of {\L}\'od\'z \\
Pomorska 149/153, 90 - 236 {\L}\'od\'z/Poland.}
 \date{}
 \begin{document}
 \maketitle  
\begin{abstract}
We derive the general formula for Lorentz - transformed spin density matrix. It is shown that an appropriate Lorentz 
transformation can produce totally unpolarized state out of pure one. Further properties, as depurification by an arbitrary
Lorentz boost and its relation to the localization properties are also discussed.
\end{abstract} \newpage
It has been shown that special relativity imposes severe restrictions on quantum information processing (for a review, see \cite{b1}).
In fact, the number of properties important from this point of view should be reexamined if relativistic corrections become
important. This can be easily seen by studying the simplest case of free relativistic particle with spin. It appeared that the
reduced density matrix for its spin is not covariant under Lorentz transformations and spin entropy ceases to be the relativistic
scalar \cite{b2}, \cite{b3}. The decrease in spin purity caused by Lorentz transformations is related to the spatial localization of the 
wave packet observed in the rest frame and any localized pure state with separate spin and momentum in the rest frame becomes
mixed when observed from moving inertial frame \cite{b4}.

The modifications enforced by special relativity are even more drastic for two- and many-particle systems \cite{b5},
\cite{b6}, \cite{b7}, both in massive and massless cases. An important property here is that the entaglement between
momenta can be transformed by Lorentz transformation to spin and vice-versa; only the joint entaglement is a Lorentz invariant notion.

The properties of relativistic transformations suggest that the protocol for quantum communication should be appropriately changed.
In this note we consider again the single particle with spin one-half. The general formula for Lorentz-transformed Bloch
vector is derived. Then, three applications are presented. First, we rederive the following result \cite{b4}: any boost applied 
to the pure spin state gives a mixed state. Second, the explicit example of general relation between depurification and localization
\cite{b4} is given. Finally, we prove that the pure state can be transformed to the state arbitrarily close to the 
totally depolarized one.

Consider quantum relativistic particle of positive mass $m$. The space of states carries an unitary irreducible representation 
of Poincare group and is spanned by $\mid p,\sigma >$- the common eigenvectors of fourmomentum and fixed (say, third)
 component of spin in the rest frame. Their scalar product reads
\begin{eqnarray}
<p',\sigma'\mid p,\sigma >=2p_0\delta^{(3)}(\vec{p}-\vec{p}')\delta_{\sigma \sigma '}\label{w1}
\end{eqnarray}
and $p^2=m^2$. An arbitrary state can be expanded in terms of basic vectors as follows
\begin{eqnarray}
\mid \psi >=\sum\limits_{\sigma}\int \frac{d^3\vec{p}}{2p_0}a(p,\sigma )\mid p,\sigma > \label{w2}
\end{eqnarray}
where
\begin{eqnarray}
a(p,\sigma )\equiv <p,\sigma \mid \psi > \label{w3}
\end{eqnarray}
is the corresponding wave function. The scalar product reads
\begin{eqnarray}
<\psi '\mid \psi >=\sum\limits_{\sigma}\int \frac{d^3 \vec{p}}{2p_0}\overline{a'(p,\sigma )}a(p,\sigma ) \label{w4}
\end{eqnarray}
Unitary irreducible representation of the Poincare group (more precisely - its universal covering $ISL(2,{\bf C})$) is
given by ($A\in SL(2,{\bf C})$)
\begin{eqnarray}
U(A,\vec{0})\mid p,\sigma >=\sum\limits_{\sigma'}D_{\sigma'\sigma}(w(p,A))\mid \Lambda p,\sigma'>;\label{w5}
\end{eqnarray}
here $\Lambda =\Lambda (A)$\ is the Lorentz matrix corresponding to $A$, $D$-unitary irreducible representation of $SU(2)$\
group while $w(p,A)\in SU(2)$\ is the so-called Winger matrix defined by
\begin{eqnarray}
w(p,A)=b^{-1}(\Lambda p)Ab(p)\label{w6}
\end{eqnarray}
Moreover, $b(p)$\ is the standard boost \cite{b8}
\begin{eqnarray}
b(p)=\frac{m+p_0+\vec{p}\cdot \vec{\sigma}}{\sqrt{2m(m+p_0)}}\label{w7}
\end{eqnarray}
The counterpart of eq. (\ref{w5}) for the wave functions reads
\begin{eqnarray}
a'(p,\sigma )=\sum\limits_{\sigma'}D_{\sigma \sigma'}(w(\Lambda^{-1}p,A))a(\Lambda^{-1}p,\sigma')\label{w8}
\end{eqnarray}
Note that for the case of spin one-half, $D(w(p,A))=w(p,A)$.

Wigner matrix has the following properties \cite{b8} \\
 (i) if $A\in SU(2),\;A^+=A^{-1}$, the corresponding Lorentz transformation is simply a rotation and 
\begin{eqnarray}
w(p,A)=A\label{w9}
\end{eqnarray}
 (ii) if $A\in SL(2,{\bf C})$\ represents pure boost corresponding to the fourvelocity $u^{\mu}=(ch\beta ,sh\beta \vec{n})$,
i.e.\begin{eqnarray}
 A=A^{+}=exp\left(-\frac{\beta}{2}\vec{n}\cdot \vec{\sigma}\right)\label{w10}
\end{eqnarray}
the relevant Wigner element can be computed to be
\begin{eqnarray}
&&w(p,A)=exp\left(i\frac{\Omega }{2}\vec{e}\cdot \vec{\sigma}\right)=cos\left(\frac{\Omega}{2}\right)+isin\left(\frac{\Omega}{2}\right)\vec{e}\cdot \vec{\sigma}\nonumber\\
&&\vec{e}\equiv \frac{\vec{p}\times \vec{n}}{\mid \vec{p}\times \vec{n}\mid}\nonumber \\
&&sin\Omega=\frac{2((1+u^0)(p^0+m)-\vec{u}\cdot \vec{p})\mid \vec{p}\times \vec{u}\mid}{((1+u^0)(p^0+m)-\vec{u}\cdot
\vec{p})^2+\mid \vec{p}\times \vec{u}\mid^2}\label{w11} \\
&&cos\Omega=\frac{((1+u^0)(p^0+m)-\vec{u}\cdot \vec{p})^2-\mid \vec{p}\times \vec{u}\mid^2}{((1+u^0)(p^0+m)-\vec{u}\cdot
\vec{p})^2+\mid \vec{p}\times \vec{u}\mid^2}\nonumber
\end{eqnarray}
Eqs. (\ref{w11}) can be, equivalently, summarized as follows
\begin{eqnarray}
w(p,A)=\frac{(1+u^0)(p^0+m)-\vec{u}\cdot \vec{p}+i(\vec{p}\times \vec{u})\cdot \vec{\sigma}}{(2(1+u^0)(p^0+m)
(m+up))^{\frac{1}{2}}}\label{w12}
\end{eqnarray}

Let us now consider the relativistic spin $-\frac{1}{2}$\ particle. The spin density matrix can be written in terms of Bloch
vector $\vec{\mu}$, 
\begin{eqnarray}
\rho=\frac{1}{2}({\bf 1}+\vec{\mu}\cdot \vec{\sigma}),\;\:\:\vec{\mu}=Tr(\rho \vec{\sigma})\label{w13}
\end{eqnarray}
The case $\mid\vec{\mu}\mid =1$\ corresponds to pure state while $\vec{\mu}=0$\ describes maximally disordered one.

Consider the pure state of relativistic particle described by the wave function $a(p,\sigma )$. The reduced density 
matrix reads
\begin{eqnarray}
\rho_{\sigma \sigma'}=\int \frac{d^3\vec{p}}{2p_0}a(p,\sigma )\overline{a(p,\sigma ')} \label{w14}
\end{eqnarray}
while the Bloch vector is given by
\begin{eqnarray}
\vec{\mu }=\int \frac{d^3\vec{p}}{2p_0}a^+(p)\vec{\sigma}a(p)\label{w15}
\end{eqnarray}
where we denoted
\begin{eqnarray}
a(p)={a(p,1)\choose a(p,2)}\label{w16}
\end{eqnarray}

Let us now apply Lorentz transformation and calculate the corresponding reduced density matrix. To this end it is sufficient
to find the new Bloch vector
\begin{eqnarray}
\vec{\mu }(\beta )=\int \frac{d^3\vec{p}}{2p_0}a^+(p)w^+(p,A)\vec{\sigma}w(p,A)a(p)\label{w17}
\end{eqnarray}

Recalling the properties (i), (ii) of Winger matrix $w(p,A)$\ we easily conclude that $\vec{\mu}$\ transforms as standard 
threevector under rotations while the identity
\begin{eqnarray}
&&w^+(p,A)\vec{\sigma}w(p,A)\equiv exp\left( -\frac{i}{2}\Omega \vec{e}\cdot \vec{\sigma}\right) \vec{\sigma}exp
\left(\frac{i}{2}\Omega \vec{e}\cdot \vec{\sigma}\right)=\nonumber \\
&&=cos \Omega\cdot \vec{\sigma} +sin\Omega (\vec{\sigma}\times \vec{e})+(1-cos\Omega )(\vec{e}\cdot \vec{\sigma})\vec{e}\label{w18}
\end{eqnarray}
gives 
\begin{eqnarray}
&&\vec{\mu}(\beta )=\int \frac{d^3\vec{p}}{2p_0}\left( cos\Omega a^+(p)\vec{\sigma }a(p)+sin\Omega (a^+(p)\vec{\sigma}a(p)
\times \vec{e})+\right. \nonumber \\
&&\left. +(1-cos\Omega )(\vec{e}\cdot a^+(p)\vec{\sigma}a(p))\vec{e}\right)\label{w19}
\end{eqnarray}
with $\vec{e}$\ and $\Omega$\ defined in eq. (\ref{w11}).

Eq. (\ref{w19}) provides the general expression for Lorentz - transformed Bloch vector. It looks rather complicated but 
appears to be quite useful. However, before making use of eq. (\ref{w19}) we shall use eq. (\ref{w17}) to show that any
Lorentz boost causes depurification \cite{b4}. Assume the reduced spin density matrix in the rest frame describes pure spin state.
Using rotation invariance to put the polarization vector in the direction of third axis we can write 
\begin{eqnarray}
&&\vec{\mu}=\vec{e}_3=\left(\int \frac{d^3\vec{p}}{2p_0}a^+(p)a(p)\right)\vec{e}_3 \label{w20} \\
&&\vec{\mu}=\vec{e}_3=\left(\int \frac{d^3\vec{p}}{2p_0}a^+(p)\sigma_3a(p)\right)\vec{e}_3 \label{w21} 
\end{eqnarray}
so that $a(p,2)\equiv 0$. Now, eq. (\ref{w17}) can be rewritten as
\begin{eqnarray}
\vec{\mu}(\beta )=\int \frac{d^3\vec{p}}{2p_0}\mid a(p,1)\mid^2\vec{e}(p)\label{w22}
\end{eqnarray}
where $\vec{e}(p)$\ is the unit vector in the direction of $a^+(p)w^+(p,A)\vec{\sigma}w(p,A)a(p)$. Multiplying both sides
of (\ref{w22}) by $\vec{\mu}(\beta )$\ and using $\int \frac{d^3 \vec{p}}{2p_0}\mid a(p,1)\mid^2=1$\ we get
\begin{eqnarray}
1-\vec{\mu}^2(\beta )=\int \frac{d^3\vec{p}}{2p_0}\mid a(p,1)\mid^2(1-\vec{\mu}(\beta )\cdot \vec{e}(p))\label{w23}
\end{eqnarray}
The necessary condition for $\mid \vec{\mu}(\beta )\mid =1$\ is that there exists a subset of positive measure such
that $u^+(p)w^+(p,A)\vec{\sigma}w(p,A)a(p)$\ has a constant direction if $\vec{p}$\ belongs to this subset. Using (\ref{w12}) 
and $a(p,2)=0$\ one easily concludes that this leads to quadratic relation between components of $\vec{p}$, contrary to 
the assumption that the measure is positive.

Let us now analyse eq. (\ref{w19}). Assume the initial state to have pure spin density matrix. Therefore, we can arrange 
things so that $a(p,2)\equiv0$. Assume, further, that $\vec{n}=\vec{e}_3$; then the last term on the RHS of eq. (\ref{w19})
vanishes. Moreover, $\Omega (p,A)$\ is axially symmetric. If we take $\mid a(p,1)\mid^2$\ to be axially symmetric, the second 
term also vanishes and
\begin{eqnarray}
\vec{\mu}(\beta )=\left(\int\frac{d^3p}{2p_0}cos\Omega \mid a(p,1)\mid^2\right)\vec{e}_3\label{w24}
\end{eqnarray}

For ultrarelativistic observer, $\beta \rightarrow \infty$,
\begin{eqnarray}
cos\Omega =\frac{(p^0-p^3+m)^2-\vec{p}^2_{\perp }}{(p^0-p^3+m)^2+\vec{p}^2_{\perp}},\;\;\;\vec{p}_{\perp}=(p^1,p^2,0)\label{w25}
\end{eqnarray}
For $p^3\approx 0,\;\vec{p}^2_{\perp}\gg m^2$\ one obtains $cos\Omega \approx 0$. Taking $a(p,1)$\ axially symmetric and 
strongly peaked arround $p^3=0$\ and $\vec{p}^2_{\perp}=M^2\gg m^2$\ we can arrange the integral (\ref{w24}) to attain
arbitrary small values; the details can be supplied easily.

We conclude that the pure state may be recognized as totally unpolarized by an observer moving sufficiently fast.

Although the initial wave function may not seem to be very appealing, our reasoning shows that there are no a priori (i.e. 
not depending on the shape of initial wave packet) bounds on disorder produced by Lorentz transformations. 

Consider now the nonrelativistic particle resting in the initial frame:\\ $<\vec{p}>=0$\ and $\mid a(p,1)\mid^2$\ is supported in the region
$\mid \vec{p}\mid \ll m$; again we assume that $\mid a(p,1)\mid^2$\ is axially symmetric. Expanding eq. (\ref{w25}) in powers of $\frac{\vec{p}^2}{m^2}$\ we find
\begin{eqnarray}
cos\Omega \simeq 1-\frac{\vec{p}^2_{\perp}}{2m^2}\label{w26}
\end{eqnarray}
Therefore,
\begin{eqnarray}
&&\vec{\mu}(\beta =\infty )=\left(1-\frac{1}{2m^2}\int \frac{d^3 \vec{p}}{2p_0}(\vec{p}_{\perp}^2\mid a(p,1)\mid^2)\right)\vec{e}_3=
 \nonumber \\
&&=\left(1-\frac{\vec{p}^2_{\perp}>}{2m^2}\right)\vec{e}_3\label{w27}
\end{eqnarray}

The mean momentum vanishes and $<\vec{p}_{\perp}^2>$\ represents the uncertainty of transverse momentum. However, in the rest 
frame everything is nonrelativistic, so all notions of nonrelativistic quantum mechanics, including that of position operator,
are well-defined. Using Heisenberg uncertainty relations one can rewrite (\ref{w27}) as
\begin{eqnarray}
\mid \vec{\mu}(\beta = \infty )\mid \leq 1-\frac{1}{8m^2}\left(\frac{1}{(\Delta x_1)^2}+\frac{1}{\Delta x_2)^2}\right)
\label{w28}
\end{eqnarray}
which gives the bound on polarization, as seen by ultrarelativistic observer, in terms of localization properties
of the state (cf. \cite{b4}).

\end{document}